\documentclass[journal]{IEEEtran}

\usepackage{graphicx}
\usepackage{amssymb}
\usepackage{amsmath}
\usepackage{mathtools}
\usepackage{algorithmic}
\usepackage{algorithm}
\usepackage{soul}
\usepackage{color}
\usepackage{gensymb}
\usepackage{cite}
\usepackage{float}
\usepackage[capitalise]{cleveref}
\usepackage{dblfloatfix}

\DeclarePairedDelimiter\norm{\lVert}{\rVert}

\newcommand{\RhatN}{$\mathbb{R}^{N}$}
\newcommand{\RhatM}{$\mathbb{R}^{M}$}
\newcommand{\RhatMN}{$\mathbb{R}^{M\times N}$}
\newcommand{\PIm}{\boldsymbol{\pi}_m}
\newcommand{\pimn}{\pi_{m,n}}
\newcommand{\phim}{\boldsymbol{\phi}_m}
\newcommand{\phimn}{\phi_{m,n}}
\newcommand{\PHI}{\boldsymbol{\Phi}}

\newcommand{\am}{\mathbf{a}_{m}}

\newcommand{\Em}{\mathbf{e}_m} %\em already existed

\newcommand{\Aphi}{\mathbf{A}_{\PHI}}

\newcommand{\emn}{e_{m,n}}
\newcommand{\gtheta}{g_{\theta}(\cdot)}

\newcommand{\nth}{n^{\text{th}}}
\newcommand{\mth}{m^{\text{th}}}

\usepackage{setspace}
\usepackage[dvipsnames]{xcolor}

\newcommand{\IHfinal}[1]{\textcolor{black}{\textcolor{black}{#1}}}
\renewcommand\hl[1]{#1}

\let\llncssubparagraph\subparagraph
\let\subparagraph\paragraph
\usepackage[compact]{titlesec}
\let\subparagraph\llncssubparagraph

\titlespacing*{\subsubsection} {0pt}{1ex plus 0.25ex minus .25ex}{0ex plus .2ex}

\begin{document}
\bstctlcite{references:BSTcontrol}

\title{Learning Sub-Sampling and Signal Recovery\\ with Applications in Ultrasound Imaging}

\author{Iris~A.M.~Huijben, Bastiaan~S.~Veeling, Kees~Janse, Massimo Mischi,~\IEEEmembership{Senior Member,~IEEE,}\newline and Ruud J.G. van Sloun,~\IEEEmembership{Member,~IEEE} \thanks{© 2020 IEEE.  Personal use of this material is permitted.  Permission from IEEE must be obtained for all other uses, in any current or future media, including reprinting/republishing this material for advertising or promotional purposes, creating new collective works, for resale or redistribution to servers or lists, or reuse of any copyrighted component of this work in other works.

I.A.M. Huijben, M. Mischi, and R.J.G. van Sloun are with the department of Electrical Engineering, Eindhoven University of Technology, The Netherlands (e-mails: \{i.a.m.huijben, r.j.g.v.sloun, m.mischi\}@tue.nl). 

B.S. Veeling is with the department of Computer Science, University of Amsterdam, The Netherlands (e-mail: basveeling@gmail.com). 

K. Janse is with Philips Research, Eindhoven, The Netherlands (\mbox{e-mail:}kees.janse@philips.com).

This research was supported in part by Philips Research, and is also part of a research program Rubicon ENW 2018-3 with project number 019.183.EN.014, which is financed by the Dutch Research Council (NWO).}}

% <-this % stops a space
% \thanks{Manuscript received April 19, 2005; revised August 26, 2015.}}

% \markboth{Journal of \LaTeX\ Class Files,~Vol.~14, No.~8, August~2015}%
% {Shell \MakeLowercase{\textit{et al.}}: Bare Demo of IEEEtran.cls for IEEE Journals}

\maketitle

\begin{abstract}
Limitations on bandwidth and power consumption impose strict bounds on data rates of diagnostic imaging systems. Consequently, the design of suitable (i.e. task- and data-aware) compression and reconstruction techniques has attracted considerable attention in recent years. Compressed sensing emerged as a popular framework for sparse signal reconstruction from a small set of compressed measurements. However, typical compressed sensing designs measure a (non)linearly weighted combination of all input signal elements, which poses practical challenges. These designs are also not necessarily task-optimal. In addition, real-time recovery is hampered by the iterative and time-consuming nature of sparse recovery algorithms. Recently, deep learning methods have shown promise for fast recovery from compressed measurements, but the design of adequate and practical sensing strategies remains a challenge. \hl{Here, we propose a deep learning solution termed Deep Probabilistic Sub-sampling (DPS), that enables joint optimization of a task-adaptive sub-sampling pattern and a subsequent neural task model in an end-to-end fashion.}
\hl{Once learned, the task-based sub-sampling patterns are fixed and straightforwardly implementable, e.g. by non-uniform analog-to-digital conversion, sparse array design, or slow-time ultrasound pulsing schemes.}
The effectiveness of \hl{our framework} is demonstrated \textit{in-silico} for sparse signal recovery from partial Fourier measurements, and \textit{in-vivo} for both anatomical image and tissue-motion (Doppler) reconstruction from sub-sampled medical ultrasound imaging data.

\end{abstract}

%TODO add keywords

\section{Introduction}

\IEEEPARstart{A}{dvanced} medical imaging techniques require transfer and storage of large amounts of data. Due to limited bandwidth, the raw sensor data must be compressed prior to its transfer to the backend system. Data compression, under-sampling, and subsequent reconstruction techniques have been an active area of research for medical imaging modalities such as computed tomography (CT) imaging \cite{sidky2008image,chen2008prior,choi2010compressed,tian2011low}, ultrasound CT imaging \cite{van2015compressed}, ultrasound (US) imaging \cite{lorintiu2015compressed,wagner2011xampling}, and magnetic resonance imaging (MRI) \cite{lustig2007sparse,lustig2008compressed}. \hl{By carefully selecting when (i.e. in time) or where (e.g. in pixel-coordinate, across sensors, or k-space) to sample, not only the amount of data can be reduced, but also the acquisition time, power drain, and, particularly for x-ray CT, radiation exposure can be lowered.}
In this paper, we therefore propose a framework for learning a task-driven sub-sampling and reconstruction method that permits reduction of sensor data rates, while retaining the information required to perform a given (imaging) task. 

Among diagnostic imaging options, US imaging is an increasingly used modality, owing to its portability, \mbox{cost-effectiveness}, excellent temporal resolution, minimal invasiveness, and radiation-free nature. \hl{Current US probes are increasingly compact, portable, and/or wireless}~\cite{acuson,butterfly,de2019ultrasound}. \hl{Across many applications and form factors, data transfer is therefore subject to very limited capacity channels. For in-body imaging (where compactness is a necessity) data is transferred from the transducer to the back-end system via a thin low-bandwidth catheter. Likewise, the aforementioned wireless, or usb-connected, probes are also heavily bandwidth constrained.} At the same time, emerging ultrafast 3D US imaging techniques \cite{provost20143d,tanter2014ultrafast} cause data rates to drastically grow, which in turn poses even higher demands on the probe-to-system communication and processing rates. Given these challenges, US imaging serves as an excellent candidate for evaluating the effectiveness of the framework that we will introduce. 

Commonly used techniques to reduce data rates in 2D and 3D echography applications are micro-beamforming \cite{larson19932,wildes20164} and slow-time\footnote{In US imaging a distinction is made between slow-time and fast-time: slow-time refers to a sequence of snapshots (i.e., across multiple transmit/receive events), at the pulse repetition rate, whereas fast-time refers to samples along depth. } multiplexing. 
The former compresses data from multiple (adjacent) transducer elements (i.e. channels) into a single focused line, thereby virtually reducing the number of receive channels. While effective, this impairs the attainable resolution and image quality. The latter only communicates a subset of the channel signals to the backend of the system for every slow-time transmission. This comes at the cost of reduced frame rates. 

Compressed sensing (CS) permits low data rate sensing (below the Nyquist rate) with strong signal recovery guarantees under specific conditions \cite{candes2006stable,candes2005decoding,candes2006compressive,candes2006near,eldar2012compressed}. In CS, a sparse signal $\mathbf{x}$ is to be recovered from measurements $\mathbf{y}$ that are taken at a sub-Nyquist rate through a sensing matrix $\mathbf{\Psi}$: $\mathbf{y}=\mathbf{\Psi}\mathbf{x}$, with $\mathbf{\Psi}$:~\RhatN$\rightarrow$\RhatM, $M\ll N$. $\mathbf{\Psi}$ should preserve distance between distant signal vectors, i.e. it should satisfy the restricted isometry property (RIP) \cite{candes2008restricted,candes2005decoding}. 

Proven (RIP-compliant) designs for $\mathbf{\Psi}$ take randomly-weighted linear combinations of input vector elements \cite{candes2006compressive,eldar2012compressed}. Unfortunately, such designs often impose challenges regarding practical implementability. For example, in US imaging, sensing weighted combinations of slow-time frames would require an, often infeasible and undesirably, large temporal signal support (including past and future values), and measuring linear combinations of channel signals imposes strong connectivity challenges. Alternatively, sampling a random subset of Fourier coefficients was also shown to be RIP-compliant \cite{candes2006compressive,eldar2012compressed}. Whenever measuring in the Fourier domain is possible (e.g. in MRI), such partial Fourier measurements alleviate the above challenges. 

After sensing, sparse signal recovery in CS is typically achieved through proximal gradient schemes, such as the Iterative Shrinkage and Thresholding algorithm (ISTA) \cite{daubechies2004iterative}. Although proximal gradient schemes are effective tools for solving non-differentiable convex optimization problems, in practice, their time-consuming iterative nature makes them less suitable for real-time applications. \hl{Moreover, these methods require knowledge of an adequate sparsifying basis, which in practice is often difficult to formalize. This is particularly true for US imaging.}

Recently, a number of deep learning approaches have been proposed for fast signal or image reconstruction in CS \cite{perdios2017deep,kulkarni2016reconnet}, showing that deep neural networks can serve as a powerful alternative to conventional recovery techniques. 

Inspired by both the challenge of finding adequate context-specific sensing matrices, and the given deep learning approaches for signal recovery, we present a deep learning solution that jointly learns a probabilistic generative model to form a context- and task-based sub-sampling pattern, and a corresponding downstream task model. \hl{We refer to the generative model as Deep Probabilistic Sub-sampling (DPS).}
Efficient \hl{joint} learning \hl{of both models} by error backpropagation is enabled through the adoption of the Gumbel-Softmax distribution \cite{jang2017categorical,maddison2016concrete,Huijben2020Deep}, that circumvents the inherently non-differentiable nature of sampling. We demonstrate \hl{our framework} for both reconstruction from partial Fourier measurements, and \hl{B-mode and tissue-motion (color Doppler) estimation} from sub-sampled \textit{in-vivo} US radio-frequency (RF) data. 

The remainder of this paper is organized as follows, we start by providing some related work in \cref{Related work}, followed by the general framework in \cref{General framework}. Section \ref{Probabilistic sampling} and \ref{TaskModel} respectively elaborate on DPS and the task model. The training strategy is described in \cref{ss:TrainingStrategy}. Section~\ref{in-silico} demonstrates our solution on a common Fourier domain sub-sampling problem. Its applications in US imaging are subsequently described in \cref{in-vivoST} and \ref{in-vivoChannels}. Results are given in \cref{Results}, which are discussed in \cref{Discussion}. Final conclusions are drawn in \cref{Conclusion}.

\hl{Throughout the paper, bold letters denote vectors (lower case) and matrices (upper case). Matrices are indexed by row-first notation, i.e. $\mathbf{a}_m$ denotes the $m^{\text{th}}$ row-vector of matrix $\mathbf{A}$, of which $a_{m,n}$ is the scalar in the $n^{\text{th}}$ column.}

\section{Related work} \label{Related work}
\noindent In this section we briefly list recent applications of conventional CS techniques for medical imaging that sub-sample the data. We then give promising applications of sparse arrays. These examples highlight the potential relevance for learning a task-driven sub-sampling pattern across a number of applications. The recent developments in deep learning for CS, that we discuss lastly, show state-of-the-art methods for learning-based data compression.

\subsection{Compressed sensing in medical imaging}
\label{sec:related_mi}

\noindent Several CS approaches have been introduced for various medical imaging applications. In MRI, CS is applied by randomly sub-sampling the K-space \cite{lustig2007sparse,lustig2008compressed}, i.e. the 2D spatial Fourier transform of the image. The authors of \cite{tsao2003k} extend this to sub-sampling in the K-time space, while preserving qualitative image reconstructions using their k-t BLAST and k-t SENSE algorithms for one coil and multiple coils, respectively. Likewise, CS has spurred low-dose X-ray CT through image reconstruction from sub-sampled projection measurements \cite{choi2010compressed,tian2011low}. 
For US imaging, in \cite{wagner2011xampling,chernyakova2014fourier}, the authors leverage CS through Xampling, by passing the RF channel signals through analog sum-of-sinc filters, thereby permitting sampling of a partial set of Fourier coefficients. Related to this, we demonstrate how \hl{DPS} permits learning of partial Fourier coefficients in \cref{ResultsA}. 
In the discrete domain, the authors of \cite{besson2016compressed_1} exploit compressed beamforming for image reconstruction from a randomly sub-sampled set of receive transducer elements, and the work in \cite{lorintiu2015compressed} proposes learned dictionaries for improved CS-based reconstruction from sub-sampled RF lines. In our work, we learn optimized discrete sub-sampling schemes from data in an end-to-end fashion.

\subsection{Sparse arrays}
\label{sec:related_sa}
\noindent Significant research efforts have been invested in exploration of adequate sparse array designs \cite{liu2017maximally}. Examples in medical US imaging are a non-uniform slow-time transmission scheme for spectral Doppler \cite{cohen2018sparse} and sparse arrays for reduction of the required number of channels for B-mode imaging\footnote{In US imaging, B-mode refers to ``brightness mode'', a 2D intensity image at a single point in time.}, based on sparse periodic arrays \cite{austeng2002sparse} or sum coarrays \cite{cohen2018beamforming}.  In \cref{ScenarioST} and \ref{ScenarioChannels}, we show how \hl{DPS} enables learning of these slow-time and array sampling patterns for US imaging in a task-based fashion. 

\subsection{Deep learning for compressed sensing}
\noindent Recently, a number of deep learning approaches have been proposed for fast signal or image reconstruction in CS \cite{perdios2017deep,kulkarni2016reconnet}, showing that deep neural networks can serve as powerful signal or image recovery methods. The authors of \cite{perdios2017deep,mousavi2015deep,mousavi2017learning,adler2016compressed,adler2016deep,lu2018convcsnet} extend learning beyond signal recovery, and simultaneously train signal compression methods. However, they all rely on taking (randomly weighted) (non)linear combinations of elements from the input vector, making them challenging to implement in hardware. Instead, \hl{DPS} is based on sub-sampling, which is straightforwardly implementable and applicable across the applications given in \cref{sec:related_mi} and \ref{sec:related_sa}.

\section{Methods} \label{Methods}

\subsection{General framework}\label{General framework}
\noindent We consider a signal vector $\mathbf{x}\in\mathbb{C}^N$ that we wish to sub-sample through binary sub-sampling matrix \mbox{$\Aphi\in\{0,1\}^{M\times N}$} parametrized by $\PHI \in$ \hl{\RhatMN}, to yield a measurement vector\footnote{$\mathbf{x}$ and $\mathbf{y}$ can also be higher dimensional. In that case all given formulas are applied on the dimension in which we want to sub-sample $\mathbf{x}$.} $\mathbf{y}\in \mathbb{C}^M$ , with $M\ll N$:
\begin{align}
\label{eqn:SubSampLASSY}
\mathbf{y} &=  \Aphi\mathbf{x}.
\end{align}
We subsequently aim to decode $\mathbf{y}$ into $\mathbf{z}$, some function of the original signal vector $\mathbf{x}$ in which we are interested (i.e. the task):
\begin{align}
    \mathbf{z} &= f(\mathbf{x}).
\end{align}
To this end, we adopt a (potentially nonlinear) differentiable function approximator $\gtheta$ parametrized by a set $\theta$:
\begin{align}
    \hat{\mathbf{z}} &= g_{\theta}(\mathbf{y}),
\end{align}
where $\hat{\mathbf{z}}$ denotes the approximation of $\mathbf{z}$ from the sub-sampled measurements $\mathbf{y}$. The function $\gtheta$ may for instance be a neural network. 
Matrix $\Aphi$ is constrained to have a row-wise $\ell_0$ norm equal to 1, i.e. every row contains exactly one non-zero element, and a column-wise $\ell_0$ norm that is either 0 or 1. As such, $\Aphi$ selects a subset of $M$ (out of $N$) elements from input vector $\mathbf{x}$. 

To permit joint learning of an adequate sub-sampling pattern for $\mathbf{x}$ and recovery of $\mathbf{z}$ through $g_\theta(\cdot)$ by backpropagation, we will introduce a probabilistic sampling strategy, on which we elaborate in the next section. 

\subsection{\hl{DPS: Deep Probabilistic Sub-sampling}}\label{Probabilistic sampling}
\noindent Each row $\am$ of $\Aphi$, with $m\in\{1,..,M\}$, is defined as a one-hot encoding\footnote{The one-hot encoding of a scalar is defined as a binary vector with exactly one nonzero value, positioned at the index given by the scalar. We define operator $\mathrm{one\_hot}_N(\cdot)$, which returns a one-hot unit-vector of length $N$.} of an independent categorical random variable
\begin{equation}
   \label{eqn:categorical}
   r_m~\sim~\textrm{Cat}(N,\PIm), 
\end{equation}
where $\PIm\in$ \RhatN~$=\{\pi_{m,1},\ldots,\pi_{m,N}\}$ is a vector containing $N$ class probabilities. Note that $\pi_{m,n}$ thus represents the probability of sampling the $\nth$ entry in $\mathbf{x}$ at the $\mth$ measurement $y_m$.
We reparametrize $\pimn$ using unnormalized log-probabilities (logits) $\phimn$, such that

\begin{equation}
\label{eqn:pimnDef}
\pimn=\frac{\mathrm{exp}~\phi_{m,n}}{\sum_{i=1}^{N} \mathrm{exp}~\phi_{m,i}},
\end{equation}
where $\phi_{m,n}$ is the $\nth$ unnormalized logit of $r_m$.

To enable sampling from the categorical probability distribution, we leverage the Gumbel-max trick \cite{gumbel1954statistical}, i.e. sampling is reparametrized into a function of the distribution parameters and a Gumbel noise vector $\Em\in$ \RhatN, with $e_{m,n} \sim \mathrm{Gumbel}(0,1)$, $n \in \{1,\ldots,N\}$ i.i.d.. For $m \in \{1,\ldots,M\}$, a realization of $r_m$ is then defined as:
\begin{equation}
\label{eq:Gumbel-Max}
     \tilde{r}_m = \underset{n \in \{1,\ldots,N\}}{\mathrm{argmax}}\big\{w_{m-1,n}+\phi_{m,n} + e_{m,n}\big\}.
\end{equation}

\hl{Previous realizations $\tilde{r}_1, \cdots, \tilde{r}_{m-1}$ are masked with $w_{m-1,n} \in \{-\infty,0\}$ for $n \in \{1\ldots N\}$, by adding $-\infty$ to the logit of the previously selected category, enforcing sampling without replacement among the $M$ distributions.}

Each row $\am \in \{1,\ldots,M\}$ can now be defined as: 
\begin{align}
    \label{eqn:onehotSample}
    &\am = \mathrm{one\_hot}_N\big\{\tilde{r}_m \big\} = \nonumber \\
    &\mathrm{one\_hot}_N\Big\{\underset{n \in \{1,..,N\}}{\mathrm{argmax}}\big\{w_{m-1,n}+\phimn+\emn\big\}\Big\}.
\end{align}        

We define $\phim \in$ \RhatN $=\{\phi_{m,1},\ldots,\phi_{m,N}\}$ as the $\mth$ row of a trainable matrix $\PHI \in$ \RhatMN that contains the unnormalized logits of all distributions. To permit optimization of $\PHI$ by backpropagation, we require $\nabla_{\phim}\mathbf{\am}$ to exist $\forall m \in \{1,\ldots,M\}$. Since $\mathrm{argmax}(\cdot)$ is a non-differentiable operator, we adopt the Straight-Through Gumbel Estimator~\cite{jang2017categorical,maddison2016concrete} as a surrogate for $\nabla_{\phim}\mathbf{\am}$:
\begin{align}
    \label{eqn:gradAm}
    &\nabla_{\phim}\am \coloneqq \nonumber \\  
    &\nabla_{\phim} \mathbb{E}_{\Em}\big[\mathrm{softmax}_{\tau}(\mathbf{w}_{m-1} +\phim+\Em)\big] = \nonumber \\
    &\nabla_{\phim} \mathbb{E}_{\Em}\Bigg[\frac{\mathrm{exp}\{(\mathbf{w}_{m-1} + \phim + \Em)\mathrm{/}\tau\}}{\sum_{i=1}^{N}\mathrm{exp}\{(w_{m-1,i} + \phi_{m,i} + e_{m,i})\mathrm{/}\tau\}}\Bigg],
\end{align}
with (row operator) $\mathrm{softmax}_{\tau}(\cdot)$ as a continuous differentiable approximation of the one-hot encoded $\mathrm{argmax}(\cdot)$ operation. We refer to sampling using the $\mathrm{softmax}_{\tau}(\cdot)$ function as soft sampling. Its temperature parameter $\tau$ serves as a gradient distributor over multiple entries (i.e. logits) in $\phim$.

In the limit of $\tau\rightarrow 0$, soft sampling approaches the one-hot encoded $\mathrm{argmax}(\cdot)$ operator in \eqref{eqn:onehotSample}~\cite{jang2017categorical,maddison2016concrete}, which results in the final trainable sub-sampling pattern of \hl{DPS}:
\begin{align}
        \label{eqn:forwardpass}
        &\am  \triangleq \lim_{\tau\rightarrow0} ~\mathrm{softmax_{\tau}}(\mathbf{w}_{m-1} + \phim + \Em), \hspace{0.1in}\text{and}\\
        \label{eqn:backwardpass}
        &\nabla_{\phim}\am \triangleq \nabla_{\phim} \mathbb{E}_{\Em}\big[\mathrm{softmax}_{\tau}(\mathbf{w}_{m-1}+\phim+\Em)\big],
\end{align}
with $\tau > 0$ and $m \in \{1,\ldots,M\}$.

\subsection{\hl{Task model}}
\label{TaskModel}
\noindent The \hl{task model} is implemented using a neural network $\gtheta$, with trainable parameters $\theta$. Suitable network architectures are application-specific and therefore described per application in ~\cref{VM}.
\subsection{Training strategy}\label{ss:TrainingStrategy}
\noindent We train model parameters $\PHI$ and $\theta$ by minimizing the mean squared error (MSE) between the model's output $\hat{\mathbf{z}}$ and the target $\mathbf{z}$, assuming normally distributed prediction errors. To prevent overfitting and exploding gradients, the problem is regularized by adding an $\ell_2$~penalty on $\theta$. Besides, we promote training towards one-hot distributions by penalizing convergence towards high entropy distributions using:
\begin{equation}
\label{eq:rowEntropy}
\mathcal{L}_{\mathcal{S}} = -\sum_{m=1}^{M}\sum_{n=1}^{N} \pimn \log \pimn,
\end{equation}
with $\pimn$ defined as in \eqref{eqn:pimnDef}.

The resulting optimization problem can be written as:
\begin{equation}
\label{eqn:NN_CSproblem}
\hat{\theta},\hat{\PHI} =  \underset{\theta,\PHI}{\mathrm{argmin}}(\mathcal{L}_\text{mse}+\mathcal{L}_\text{pen}),
\end{equation}
with
\begin{equation}
\label{eqn:LossFunc}
\mathcal{L}_\text{mse} = \mathbb{E}_{(\mathbf{x},\mathbf{z})\sim p_{_{\mathcal{D}}}}\norm{\mathbf{z}- g_{\theta}(\Aphi\mathbf{x})}_2^2,
\end{equation}
and
\begin{equation}
\label{eqn:LossPen}
\mathcal{L}_\text{pen} = \lambda\norm{\theta}_2 + \mu\mathcal{L}_{\mathcal{S}},
\end{equation}
where the input and target vectors, i.e. $\mathbf{x}$ and $\mathbf{z}$ respectively, follow data-generating distribution $p_{_{\mathcal{D}}}$. Penalty multipliers $\lambda$ and $\mu$ weigh the importance of the different penalties. 

The Adam solver with hyperparameters \mbox{$\beta_1 = 0.9$}, $\beta_2~=~0.999$, and $\epsilon=1e{-7}$ \cite{kingma2014adam} is used to stochastically optimize~\eqref{eqn:NN_CSproblem}. In practice, we found that the appropriate learning rates for $\PHI$ and $\theta$ were different. As such, two distinct learning rates were used, i.e. $\eta_{\PHI}$ and $\eta_{\theta}$, \hl{by multiplying the update step for $\PHI$ by a factor $\frac{\eta_{\PHI}}{\eta_{\theta}} > 1$}. The adopted values are reported in \cref{VM}, along with the values for the penalty multipliers $\lambda$ and $\mu$. 

The temperature parameter $\tau$ in \eqref{eqn:backwardpass} is initialized at $5.0$ and gradually lowered to $0.5$ during training. \hl{This annealing scheme ensures that soft sampling approaches hard sampling towards the end of training, while confining the variance of the gradient updates at the beginning of training} \cite{jang2017categorical}.

\hl{Logits matrix $\PHI$ was initialized to have a prior on training towards a diagonal matrix, i.e. }all elements $\phimn$, with $m \in$ $\{1,\ldots,M\}$ and $n \in \{1,\ldots,N\}$ are initialized according to:
\begin{equation}
\label{eqn:customInit}
    \hat{\phi}_{m,n} = \alpha(n-\frac{N}{M}m)^4 + \beta(n-\frac{N}{M}m)^2 + \gamma_{m,n},
\end{equation}
with constants $\alpha=-2.73e{-7}$ and $\beta=-2.73e{-3}$, and  $\gamma_{m,n}\sim \mathcal{N}(0,0.01)$ i.i.d..

\begin{figure}
\centering
\includegraphics[width=\linewidth,trim={0cm 16.6cm 9.5cm 1.5cm},clip]{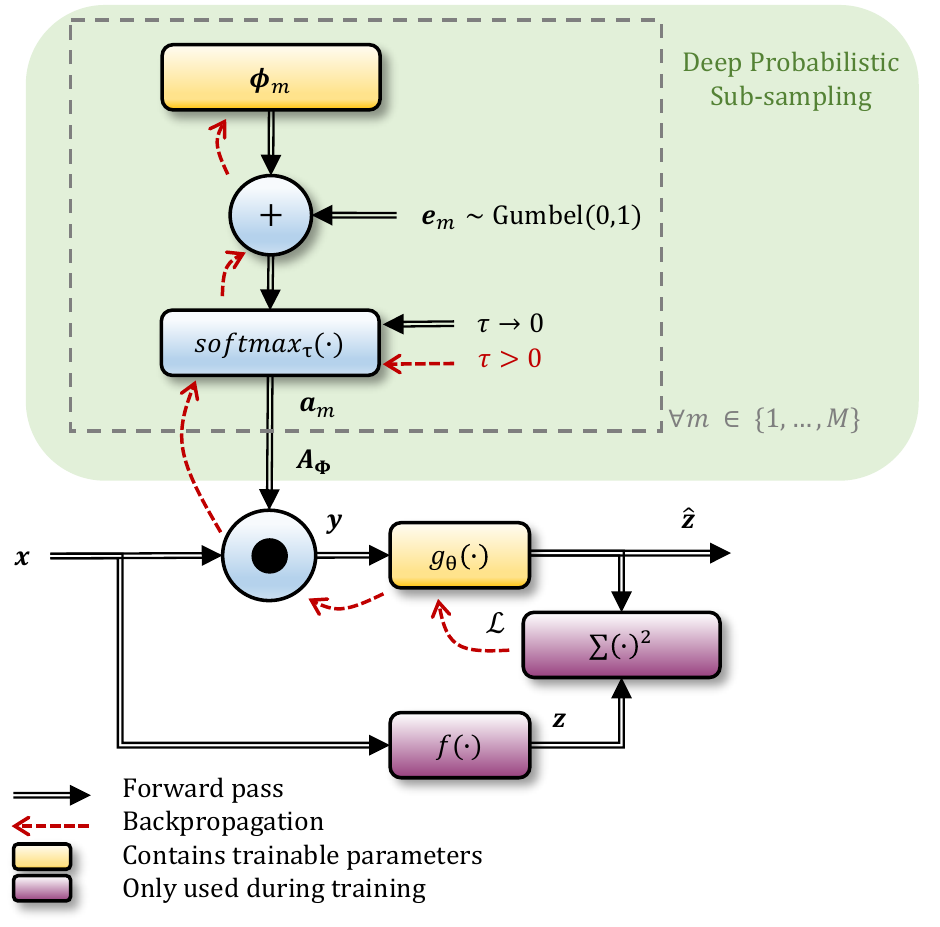}
\caption{An overview of generative sub-sampling model DPS with task model $\gtheta$, with gradient backpropagation depicted in red. The $\odot$ symbol indicates a matrix multiplication between $\Aphi$ and the dimension to be sub-sampled of $\mathbf{x}$. $\mathbf{y}$ is a subset of $\mathbf{x}$, which is used to approximate $\mathbf{z}$ by $\hat{\mathbf{z}}$. \hl{Note, for the sake of clarity we left out the masking from Algorithm}~\ref{LASSYalgorithm}.}
\label{fig:NNstructure}
\end{figure}

A system-level overview of DPS followed by the task model is visualized in~\cref{fig:NNstructure}, and its pseudocode is given in Algorithm \ref{LASSYalgorithm}. It was implemented in Python using Keras \cite{chollet2015keras} with a TensorFlow backend \cite{abadi2016tensorflow}. Training and inference were performed on a Titan XP (NVIDIA, Santa Clara, CA).
\begin{algorithm}
\caption{\hl{Deep Probabilistic Sub-sampling (DPS)}}
\label{LASSYalgorithm}
\begin{algorithmic}
\REQUIRE Training dataset $\mathcal{D}$, Number of iterations $n_{\text{iter}}$, $\tau_{\text{init}} = 5.0$, $\tau_{\text{end}} = 0.5$, $\mathbf{w}_0=\mathbf{0}$, initialized trainable parameters $\PHI$ and $\theta$.
\ENSURE Trained logits matrix $\PHI$ and reconstruction network parameters $\theta$.    

\STATE{ - Compute $\Delta\tau = \frac{\tau_{\text{init}} - \tau_{\text{end}}}{n_{\text{iter}}-1}$ \\
}
\FOR{$i=1$ to $n_{\text{iter}}$}
    \STATE{ - Draw mini-batches $\mathbf{x}_i$: a random subset of $\mathcal{D}$}
    \STATE{ - Compute fully sampled target: $\mathbf{z}_i = f(\mathbf{x}_i)$}
    \STATE{- Initialize mask: $\mathbf{w}_0 =\mathbf{0}$} \\
    \STATE{ - Compute $\Aphi = [\mathbf{a}_{1}; \ldots; \mathbf{a}_{M}]$ using:}
    \FOR{$m = 1$ to $M$}
       \STATE{- Draw i.i.d. Gumbel noise samples $\mathbf{e}_m \in \mathbb{R}^{N}$} \\
        \STATE{- Sample from the distribution: \\$\tilde{r}_m =
            \underset{n \in \{1\ldots N\}}{\mathrm{argmax}}\big\{w_{m-1,n}+\phimn+e_{m,n}\big\}$} \\
        \STATE{- $ \text{Create one-hot vector: } \mathbf{a}_m = \mathrm{one\_hot}_N(\tilde{r}_m)$}\\
        \STATE{- Take current mask: $\mathbf{w}_m = \mathbf{w}_{m-1}$}
        \STATE{- Update mask: $w_{m,\tilde{r}_m} = -\infty$}
    \ENDFOR\\
    \STATE{ - Sub-sample the signal: $\mathbf{y}_i = \Aphi\mathbf{x}_i$}
    \STATE{ - Compute reconstruction: $\hat{\mathbf{z}}_i = g_{\theta}(\mathbf{y}_i)$}
    \STATE{ - Compute loss using: $\mathcal{L}_i = \norm{\mathbf{z}_i - \hat{\mathbf{z}}_i}_2^2 + \mathcal{L}_\text{pen}$}
    \STATE{ - Set $\tau = \tau_{\text{init}} - (i-1) \cdot \Delta \tau$} \\
    \STATE{ -  Set $\nabla_{\phim} \am = \nabla_{\phim} \mathbb{E}_{\Em}\big[\mathrm{softmax}_{\tau}(\mathbf{w}_{m-1}+\phim+\Em)\big]$}
    \STATE{- Use Adam optimizer to update $\PHI$ and $\theta$}
\ENDFOR
\end{algorithmic}
\end{algorithm}

\section{Validation methodology} \label{VM}
\subsection{Partial Fourier sampling of sparse signals} \label{in-silico}
\subsubsection{Data generation and task}\label{in-silico:Data}
\noindent Many practical CS applications require signal reconstruction from partial Fourier measurements~\cite{lustig2007sparse,lustig2008compressed,otazo2010combination}, and we therefore first demonstrate \hl{DPS} in such a scenario.
To that end, we synthetically generate random K-sparse signal vectors $\mathbf{z}\in\mathbb{R}^{128}$, with $K=5$, which we subsequently Fourier-transform to yield the signal $\mathbf{x}\in\mathbb{C}^{128}$ that we aim to partially sample\footnote{For each experiment the length of the signal (in the dimension to be sub-sampled) was set to the integer multiple of the sub-sampling factor that is closest to 128, e.g. 126 for factor 6.}. \hl{Training data is generated on-line, but we create a fixed hold-out test set of 1000 of such signals for model comparison}. 

In this experiment, the measurement $\mathbf{y}\in\mathbb{C}^{M}$, with $M\leq 128$, is a sub-sampled set of Fourier coefficients in $\mathbf{x}$, and the task is to recover the sparse signal $\mathbf{z} = f(\mathbf{x}) =  \mathcal{F}^{-1}(\mathbf{x})$, from measurement $\mathbf{y}$, where $\mathcal{F}^{-1} \in \mathbb{C}^{N\times N}$ is the inverse discrete Fourier transform (IDFT).

\hl{According to CS theory, the sensing matrix should be RIP-compliant in order to have strong signal reconstruction guarantees} \cite{candes2008restricted,candes2005decoding}. \hl{Here, this sensing matrix is defined as $\Psi=\Aphi\mathcal{F}$. Since the sparse basis, i.e. $\mathcal{F}$, is known, we can check whether $\Psi$ adheres to RIP for the learned sampling matrix $\Aphi$. For our K-sparse signals, each $M \times K$ (with $M > K$) sub-matrix should be full rank, and all $M$ rows should be (close to) orthogonal to each other. Since $\Psi$ is a subset of the rows of the ortho-normal DFT matrix $\mathcal{F}$, we only need to check the rank of each of the sub-matrices.}

\subsubsection{Task model}\label{in-silico:TaskModel}
We adopt a specific \hl{task} model architecture $\gtheta$, inspired by the proximal gradient ISTA scheme~\cite{daubechies2004iterative}. It unrolls the iterative solution of ISTA as a 3-layer feedforward neural network with trainable parameters~\cite{gregor2010learning,chen2018theoretical}. To prevent dying gradients during backpropagation, we replace the conventional soft-thresholding operators in this learned ISTA (LISTA) method by a sigmoid-based soft-thresholding operator~\cite{atto2008smooth}. The thresholds are jointly trained as well.

\hl{We compare signal reconstruction by DPS to reconstruction from untrained fixed uniform or random sub-sampling patterns.} The latter is typically adopted in CS~\cite{candes2006compressive,eldar2012compressed}. \hl{The adopted task model is jointly trained with the specified sub-sampling method. We refer to these methods as \mbox{\textit{Uniform+LISTA}}, \textit{Random+LISTA}, and \textit{DPS+LISTA}, respectively. Additionally, we compare these task models to the ISTA reconstruction for (the same) random and learned sub-sampling realizations. These experiments are referred to as \textit{Random+ISTA} and \textit{DPS+ISTA}, respectively.}

\subsubsection{Training}\label{in-silico:Train}
We train for 96,000 iterations across \hl{randomly on-line generated} mini-batches of 16 Fourier-transformed data vectors. We define one iteration as a parameter update using one mini-batch of data. The learning rates $\eta_{\PHI}$ and $\eta_{\theta}$ are set at $5e{-3}$ and $1e{-3}$, and the penalty multipliers $\lambda$ and $\mu$ at 0.0 and $1e-8$, respectively.  \\*[0.1ex]

\subsection{Slow-time sub-sampling in ultrasound imaging} \label{in-vivoST}

\subsubsection{Data acquisition and pre-processing}\label{dataAcquisition}
Sequential (slow-time) echography data were acquired from \hl{two different porcine models}. To that end, a 48-element linear array miniTEE s7-3t transducer with a pitch of 0.151 mm was used in combination with a Verasonics Vantrage system (Kirkland, WA). The center frequency for transmission and reception was 4.8 MHz and a 13-angle diverging wave scheme was used (with steering angles uniformly distributed between $-45\degree$ and $+45\degree$, and $F\text{-number} = -1.91$). The sampling rate of the received RF data was 19.2 MHz and coherently compounded beamformed frames (each with 68 scanlines) were collected at a frame rate of 474 Hz. For both the training and test set, all RF data frames were demodulated into their in-phase and quadrature (IQ) components, and subsequently normalized between -1 and +1.

\hl{Data belonging to either the training or test set differ in terms of the used porcine model, viewing angle, and location of the probe. The training set contains \textit{in-vivo} open-chest extracardiac US measurements of the right atrium of the first porcine model. Two ablation surgeries took place, causing physical changes in the movement of the heart. Measurements with two different viewing angles, before, in between, and after the two ablations were included in this training set, leading to a total of 1,340 frames.

The test set contains \textit{in-vivo} open-chest extracardiac US measurements from the second porcine model. This model had an infarct two months before the acquisition. The measurements were made at the left ventricular posterior wall with the long axis view. In total, 335 frames were recorded.}

\begin{figure}[t]
\includegraphics[width=\linewidth,trim={0cm 15.4cm 9.4cm 1.2cm},clip]{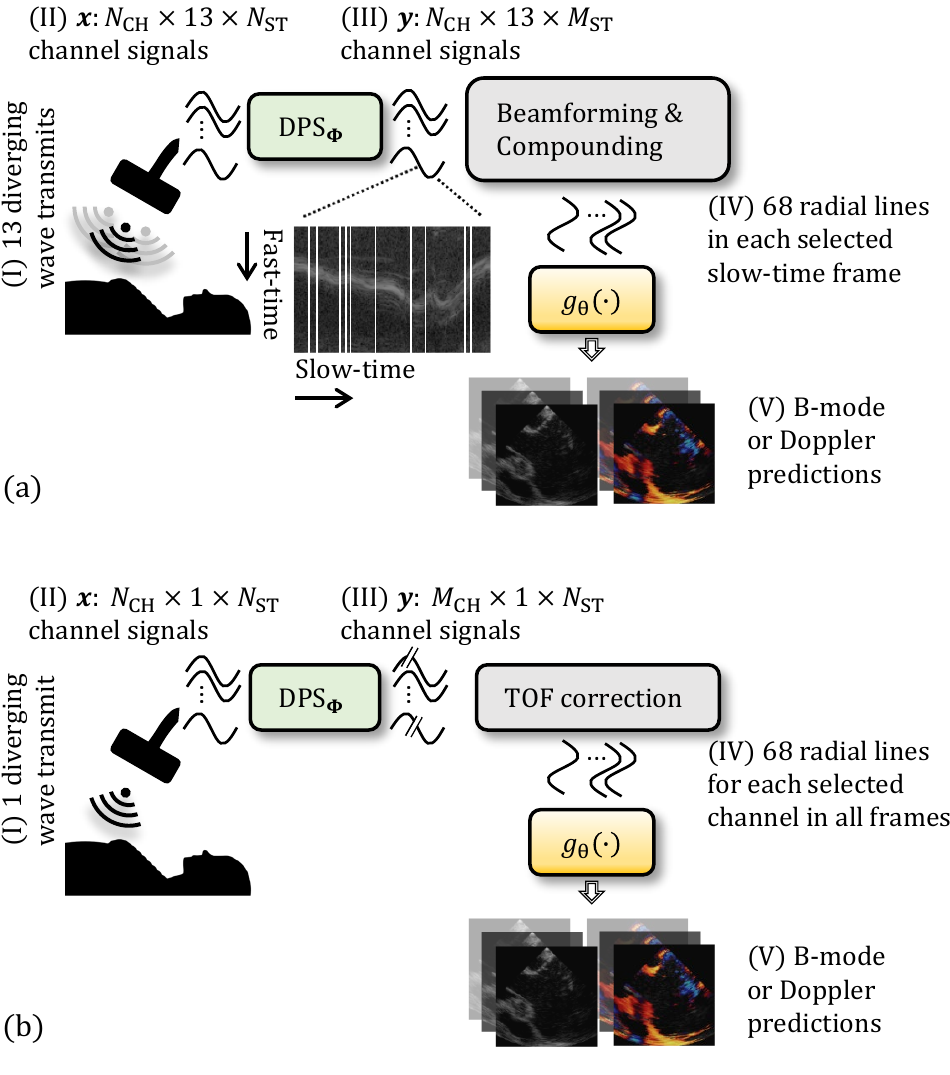}
\caption{An illustration of the sensing pipeline for the considered US experiments in this paper. The sub-sampling model (DPS) is trained jointly with the task model $\gtheta$, to (a) learn a sparse slow-time transmit scheme, selecting $M_{\text{ST}}$ from $N_{\text{ST}}$ transmit events \IHfinal{(each event includes 13 diverging wave transmits)}, and (b) learn a sparse set of channel signals, sampling  $M_{\text{CH}}$ out of $N_{\text{CH}}$ channels of the full array.}
\label{fig:data_dimensions}
\end{figure}

\hl{The experiments were carried out with prior approval from the Dutch Central Commission for Animal Studies (protocol nr. AVD/115002015205) (The Netherlands) and according to the European directives (2010/63/EU) (Belgium) and the Guidelines for the Care and Use of Laboratory Animals (National Institutes of Health, USA).}

\subsubsection{Tasks}\label{targets}
Using the training dataset acquired according to the procedure described in the previous section, we employ \hl{DPS} to learn a sparse slow-time transmit scheme\footnote{\IHfinal{DPS learns to select $M_{\text{ST}}$ out of $N_{\text{ST}}$ slow-time frames, of which each is created by coherently compounding 13 diverging wave transmits. The group of diverging waves corresponding to a single slow-time frame is thus always sampled as a whole in this scenario.}}, tailored towards two distinct imaging tasks. As a first task, we consider recovery of the envelope of the beamformed RF signal in order to produce anatomical B-mode US images. Here, $\mathbf{x}$ is (fully sampled) complex IQ data,  $\mathbf{y}$ is a subset of slow-time transmits in $\mathbf{x}$, and the target $\mathbf{z}$ is the magnitude of $\mathbf{x}$. Second, we explore learning-based tissue-motion estimation (i.e. \hl{color} Doppler \cite{van2018learning,van2019deep}). In this case, the target $\mathbf{z}$ is computed by applying the Kasai auto-correlator \cite{kasai1985real} on $\mathbf{x}$. We mask the Doppler targets such that Doppler information is only available at positions where tissue is present. We expect the two tasks to yield very distinct sampling patterns; while B-mode images are independently constructed per frame, Doppler shifts are obtained by measuring phase shifts across the slow-time sequence. \Cref{fig:data_dimensions}a illustrates the processing pipeline and associated data dimensions.

\subsubsection{Task model}\label{in-vivo_architecture}
For prediction of $\hat{\mathbf{z}}$ from the subset of slow-time transmit events in $\mathbf{y}$, we employ a convolutional neural network~\cite{Goodfellow-et-al-2016}. The architecture of this task model differs slightly for the two defined tasks. For B-mode recovery, the model comprises a total of six convolutional layers, of which the first two are 1-dimensional and applied across the fast-time dimension (with respectively 256 and 128 features, and window length = 5). The last four convolutions are 2-dimensional and applied across both fast-time and slow-time, all having kernel sizes $5\times5$ and respectively 32, 64, 32, and 1 feature(s). The receptive field of this task model across fast-time, sub-sampled slow-time, and radial scanlines was 25, $M_{\text{ST}}$ (i.e. all sampled slow-time frames), and 1, respectively.

For Doppler reconstruction, $\mathbf{y}$ is first zero-filled, i.e. zeros are added at non-sampled indices, after which six 2D convolutional layers with $5\times5$ kernels and respectively 32, 64, 128, 64, 32, and 1 feature(s) are leveraged. Applying zero-filling after sub-sampling retains the original time-index of the sampled frames, which we found beneficial for Doppler recovery. The receptive field for fast-time, sub-sampled and zero-filled slow-time, and scanlines is 25, 25, and 1, respectively.
In both task models, all convolutional layers, except the last, have leaky ReLU activations ($\alpha=0.1$) \cite{xu2015empirical}.

\hl{Next, for B-mode recovery, the model contains four 2D convolutions layers (over fast-time and slow-time) having kernel sizes $5\times5$ and respectively 32, 64, 32, and 1 feature(s). The relatively more difficult task of Doppler recovery required a slightly deeper model, containing six 2D convolutional layers with kernel sizes $5\times5$ and respectively 32, 64, 128, 64, 32, and 1 feature(s).} We use leaky rectified linear unit (leaky ReLU) activation functions ($\alpha=0.1$) across all convolution layers, except the last, which has no activation function \cite{xu2015empirical}.
% %
\subsubsection{Training}
 For both tasks, the models are stochastically optimized using the Adam solver, with settings as described in \cref{ss:TrainingStrategy}, and learning rates $\eta_{\PHI}=2e{-3}$ and $\eta_{\theta}=1e{-4}$.
We train for 640,000 iterations with mini-batches consisting of 16 randomly selected patches. Each patch contains 128 sequential slow-time samples of 256 fast-time IQ samples for a single radial scanline. The logits in $\PHI$ are initialized according to~\eqref{eqn:customInit}. Penalty multipliers $\lambda$ and $\mu$ are set at $1e{-5}$ and $1e{-7}$, respectively, \hl{and $\mu$ linearly increases with $1e{-8}$ per epoch (1 epoch consists of 64 mini-batch updates) to more strictly promote low-entropy distributions towards the end of training.}
\subsection{Channel sub-sampling in ultrasound imaging} \label{in-vivoChannels}
\subsubsection{Data acquisition and pre-processing}
\label{sec:dataForChannelSub}
The imaging setup and division into training and test set as described in \cref{dataAcquisition}, was also used to demonstrate \hl{DPS} for sub-sampling across the 48-channel array, \hl{prior to receive beamforming from one diverging wave transmit.} 
To facilitate the subsequent receive beamforming stage, we first pre-delay the channel signals for 68 different scanlines (with steering angles $\zeta$ in $[-\frac{\pi}{4},\frac{\pi}{4}]$) \cite{hasegawa2011high}. Taking into account the transmit delay $T_T$ (i.e. the time-of-flight (TOF) between the virtual point source behind the array and the focus point in our diverging wave transmission scheme), and the receive delay $T_R$ (i.e. the TOF of the back-scattered wave between the focus point and the array element location, indexed by $i$), the total delay function $T(i,\zeta)$ for the central wave transmit is defined as~\cite{hasegawa2011high}:
\begin{equation}
T(i,\zeta) = T_T(i,\zeta) + T_R(i,\zeta),
\end{equation}
in which
\begin{equation}
T_T(i,\zeta) = \frac{\sqrt{f_d^2 \sin^2\zeta - (r_f+f_d\cos\zeta)^2} - r_f}{c_0},
\end{equation}
and
\begin{equation}
T_R(i,\zeta) = \frac{\sqrt{f_d^2\cos^2\zeta + {\{f_d\sin\zeta-\mathrm{\Delta x}(i-\frac{L-1}{2})\}}^2}}{c_0}.
\end{equation}
Focal depth is denoted by $f_d$, $r_f$ is the distance between the surface of the transducer array and the virtual point source behind the array, $\Delta x$ and $L$ are respectively the pitch and total number of channels of the array, and $c_0$ denotes the speed of sound in soft tissue. The adopted values for these parameters are: $f_d=40$~mm, $r_f=13.5$~mm, ${\Delta x=0.151}$~mm, \mbox{$L=48$}, and $c_0=1540$~m/s. 

After computing 68 delayed signals per channel, we obtain a 4D dataset spanning slow-time frames, fast-time samples, channels, and radial scanlines. Note that pre-computing these delays is merely only done to accelerate training, and can in practice be performed post-sampling, as shown in \cref{fig:data_dimensions}b. Finally, the pre-delayed RF channel signals are demodulated into their IQ components, and thereafter normalized between -1 and +1.

\subsubsection{Tasks}
We again distinguish two tasks, \hl{B-mode reconstruction} and color Doppler estimation. For both tasks, targets are generated using fully sampled DAS-beamformed channel data from one wave transmit. These beamformed IQ-demodulated RF signals are processed as described in \cref{targets} to create B-mode, respectively Doppler, targets. 

\subsubsection{Task model}
For recovery of $\hat{\mathbf{z}}$ from the sub-sampled channel data in $\mathbf{y}$, we leverage a convolutional neural network having the same architecture for both tasks. \hl{After sub-sampling, we first apply zero-filling for the non-sampled channels, after which four 2D convolutional layers, having $5\times 5$ kernels and respectively 64, 128, 64, and 48 features, are adopted}. Convolutions take place across the fast- and slow-time dimension, i.e. the channels are fully connected. Each convolutional layer is followed by a leaky ReLU activation function ($\alpha=0.1$) \cite{xu2015empirical}. The network's last layer is a fully connected layer across the (sub-sampled) channel dimension, and therefore acts as an array apodization, as used in typical DAS beamforming \cite{szabo2004diagnostic}. The receptive field of this model for respectively fast-time, slow-time, sub-sampled channels, and scanlines was 17, 17, $M_{\text{CH}}$ (i.e. all selected channels), and 1.
\subsubsection{Training}
For both tasks, the networks are stochastically optimized using the Adam optimizer, with its settings as described in \cref{ss:TrainingStrategy}. The learning rates $\eta_{\PHI}$ and $\eta_{\theta}$ are set at $2e{-3}$ and $1e{-4}$ respectively, and we train for 192,000 mini-batch updates. Randomly selected mini-batches consist of 16 patches spanning 32 slow-time frames, 64 fast-time samples, 48 channels, and one radial scanline. The trainable matrix $\PHI$ is initialized according to \eqref{eqn:customInit} and hyperparameter $\lambda$ is set at $1e{-5}$. The \hl{entropy penalty multiplier $\mu$ is set at $1e{-8}$ and linearly increases with $1e{-11}$ per epoch (1 epoch consisted of 64 mini-batch updates), to more strictly promote one-hot distributions towards the end of training.} 
\begin{figure}[t]%
\includegraphics[width=\linewidth,trim={0cm 16.8cm 9.5cm 1.15cm},clip]{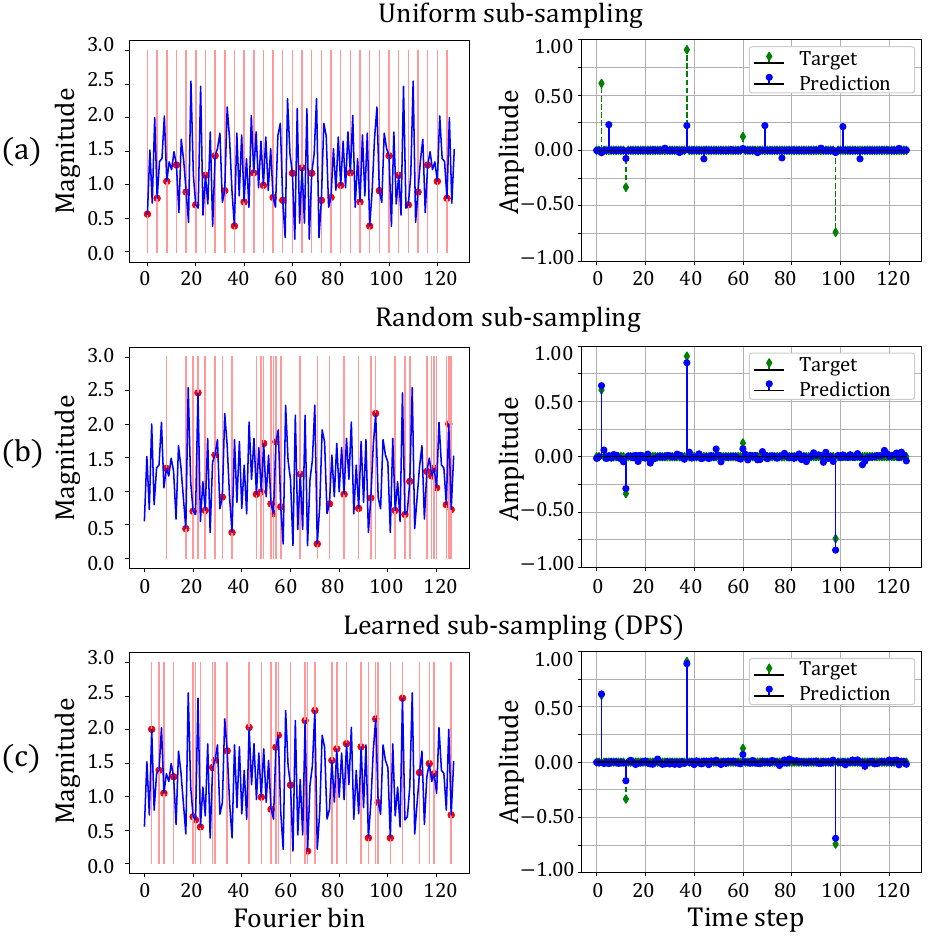} %
\caption{Left column: Fixed uniform (a), fixed random (b) and learned (using \hl{DPS}) (c) sub-sampling patterns (sub-sampling factor $\frac{N}{M}=4$), with selected samples indicated in red. Right column: Signal recovery (blue) and ground truth signal (green) of an example signal from the test set.}%
\label{fig:FourierResults}
\end{figure}
% %
\begin{figure}[t]
\centering
\includegraphics[width=\linewidth,trim={0cm 20.93cm 9.5cm 1.85cm},clip]{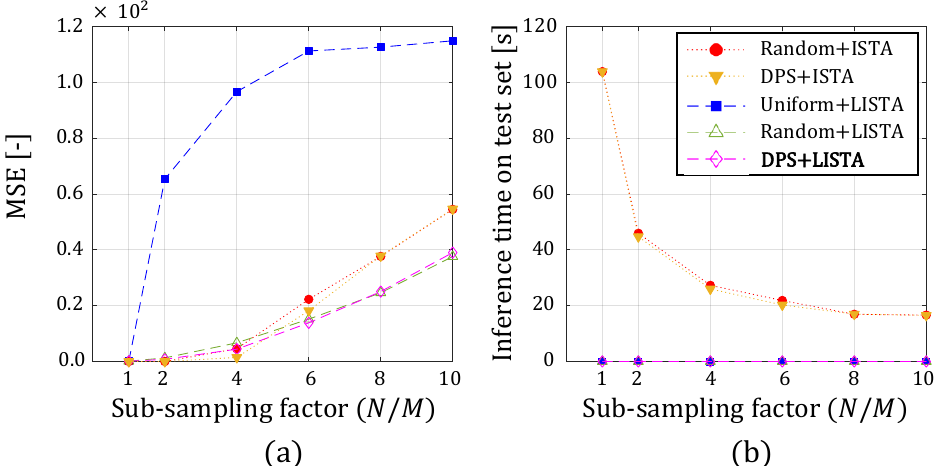}
\caption{Averaged inference results across all signals from the hold-out test set. The ISTA algorithm run for 300 iterations, while the unfolded ISTA, i.e. Learned ISTA (LISTA), was unfolded only 3 times. (a) MSE values show that uniform sub-sampling strongly hampers sparse signal reconstruction. (b) Inference time on the test set is much larger for the iterative ISTA algorithm, compared to LISTA. For ISTA, inference time also considerably grows for lower sub-sampling factors, while this effect is not visible for LISTA. The inference times for LISTA were found to be in the order of 0.01 seconds.} 
\label{fig:MSEgraphFourier}
\end{figure}

\section{Results}\label{Results}
\subsection{Partial Fourier sampling of sparse signals}\label{ResultsA}
\noindent \Cref{fig:FourierResults} displays sparse signal recovery from partial Fourier measurements for an example signal from the test set. \hl{The Fourier space is sub-sampled (with factor $\frac{N}{M}=4$) uniformly (a), randomly (b), or using a realization from the learned sub-sampling distributions by DPS (c)}. 
A quantitative evaluation of the signal recoveries for different sub-sampling and reconstruction methods is given in \cref{fig:MSEgraphFourier}a.  Uniform sub-sampling performed poorly due to aliasing, resulting in a repeated prediction pattern (\cref{fig:FourierResults}a-right). However, (CS-inspired) random sampling (\cref{fig:FourierResults}b-right) shows on par results with learned sampling by \hl{DPS} (\cref{fig:FourierResults}c-right). Interestingly, the learned pattern also exhibits (pseudo-random) irregular sampling (\cref{fig:FourierResults}c-left), and its corresponding sensing matrix $\PHI$ showed to be RIP-compliant. 

\hl{Comparing signal reconstruction using either ISTA (300 iterations) or LISTA (3 folds),} \cref{fig:MSEgraphFourier}a \hl{shows that for high sub-sampling factors, LISTA outperforms ISTA, while enabling much faster (speed-up factor $>1000$) inference} (\cref{fig:MSEgraphFourier}b).\hl{ This finding confirms the results provided by the authors of} \cite{chen2018theoretical}, \hl{who show that an unfolded scheme can achieve better performance for few unfoldings compared to the iterative scheme for the same amount of iterations.} 

\begin{figure*}[t]
\includegraphics[width=\linewidth,trim={0cm 14.35cm 0.3cm 0cm},clip]{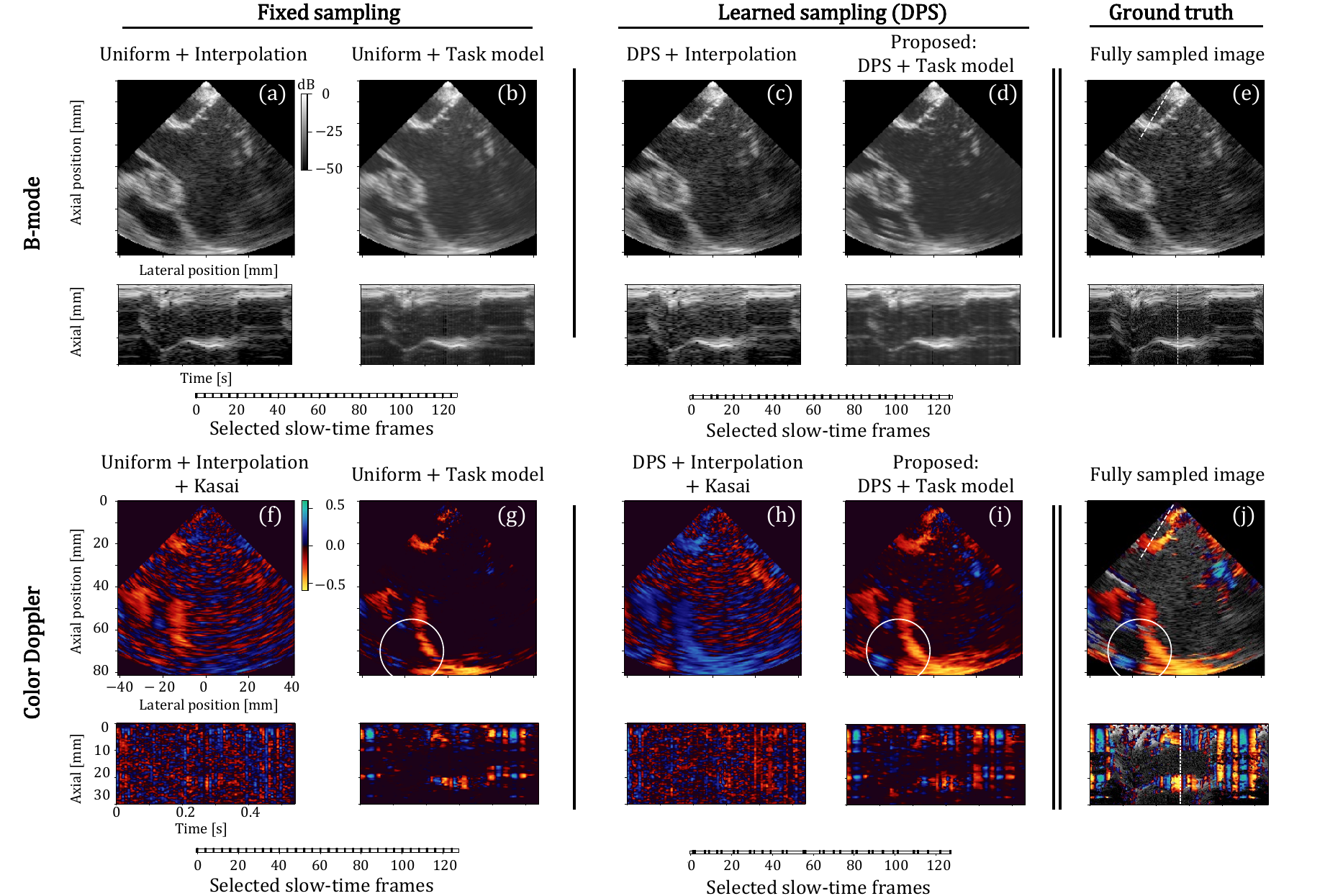}
\caption{Reconstructed anatomical B- and M-mode (a-e) and color Doppler images (f-j) after sub-sampling by a factor 4 across slow-time frames, either uniformly (a, b, f, g), or learned by DPS (c, d, h, i). For both sampling strategies we compare the jointly learned task model $\gtheta$ (b, g, d, i) to a naive non-learned method for the task at hand (cubic interpolation in a and c, followed by the Kasai auto-correlator in f and h). (e, j) Fully sampled images as a reference. \IHfinal{For fair comparison, we display a frame that is not sampled by either of the compared sub-sampling strategies (a-i).} The dashed lines in the top and bottom image indicate the selected radial M-mode line and B-mode frame, respectively.}
\label{fig:SlowTimeresults}
\end{figure*}
\begin{figure}[t] 
\includegraphics[width=\linewidth,trim={0cm 20.7cm 9.4cm 1.4cm},clip]{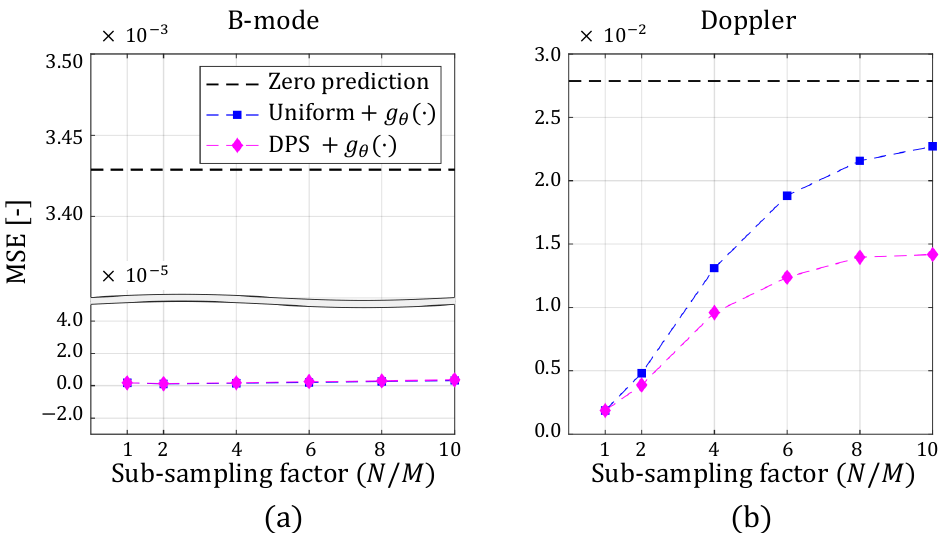}
\caption{MSE for B-mode (a) and Doppler (b) recovery from a sub-set of slow-time transmits, averaged over the hold-out test set. The black dashed lines indicate the MSE in case only zero values are predicted.}
\label{fig:MSEgraphST}
\end{figure}
\subsection{Slow-time sub-sampling in ultrasound imaging}
\label{ScenarioST}
\noindent \Cref{fig:SlowTimeresults} demonstrates anatomical B-mode (a-e) and color Doppler (f-j) reconstructions for both fixed uniform and DPS-driven sub-sampled slow-time transmits. For uniform sub-sampling, the DPS block in \cref{fig:data_dimensions}a is replaced by a (non-trainable) uniform sub-sampling operation. The adopted sampling patterns (with selected frames in black) are visualized below the reconstructions. Note that for DPS, these sampling patterns correspond to a single stochastic realization from the trained probability distributions, on which we detail in Appendix~\ref{App:probs_DPS}. To demonstrate both the individual and joint contributions of DPS and the task-model to performance, we moreover compare our trained neural network task model to a non-learned cubic-interpolation-based reconstruction method for both sampling strategies.

Interestingly, \hl{DPS}' learned sampling patterns for both tasks showed to be very distinct. For B-mode reconstruction, the learned pattern exhibited an almost perfectly uniform sampling pattern (\cref{fig:SlowTimeresults}c and d), whereas For Doppler reconstruction, the sub-sampling pattern found by DPS exhibited an `ensemble'-type of sampling (\cref{fig:SlowTimeresults}h and i). For both tasks, similar patterns were clearly visible for the other tested sub-sampling factors as well. The learned `ensemble'-style sub-sampling pattern for Doppler recovery efficiently captures high frequency slow-time signals due to tissue displacements (Doppler shifts) within ensembles, and relatively low frequency information (changes in Doppler shifts over time) among these ensembles. Given the (nearly) uniform sampling pattern learned by DPS for B-mode reconstruction, the resulting reconstructions using fixed uniform versus learned sub-sampling (\cref{fig:SlowTimeresults}a, b and \ref{fig:SlowTimeresults}c, d, respectively), were found to be similar. For increasing factors, reconstructed images displayed increased blurring for all sampling-reconstruction combinations. The MSEs for different sub-sampling factors, both for uniform and learned sampling by DPS combined with task model $\gtheta$, are visualized in~\cref{fig:MSEgraphST}a. It confirms the similarity of reconstructed B-mode images for uniform and learned sub-sampling patterns.

The need for accompanying DPS with a jointly trained nonlinear task model becomes particularly evident when comparing the Doppler reconstructions in \cref{fig:SlowTimeresults}h and i. Simply interpolating frames and using a standard Doppler estimator (Kasai) leads to significant aliasing in \cref{fig:SlowTimeresults}h. For anatomical B-mode imaging, which does not rely on inter-frame processing, the difference between reconstruction methods mainly manifests itself by the inherent denoising priors of the chosen convolutional architecture.

That joint training of sampling and a dedicated task model yields best results can also be seen in \cref{fig:MSEgraphST}b, where we quantitatively compare fixed versus learned sampling.

Using the trained network for inference on 256 slow-time frames, containing 68 scanlines and 2048 fast-time samples from the test set, took on average 1.29 s ($\sigma=29.2$~ms). Accordingly, the reconstruction network allows a reconstruction speed of 198 sub-sampled frames per second.

\begin{figure*}[t]
\includegraphics[width=\linewidth,trim={0cm 13.4cm 0.5cm 0cm},clip]{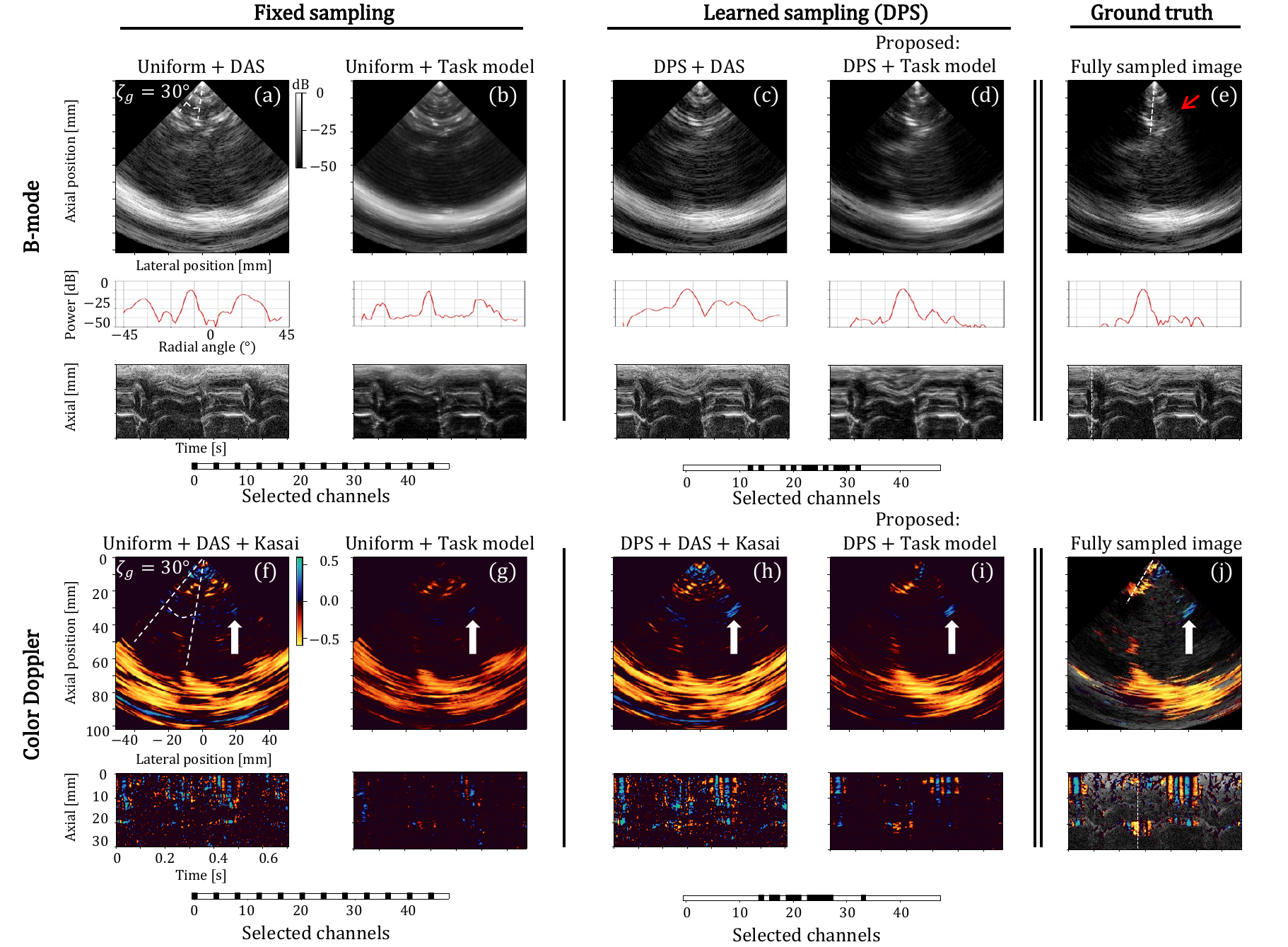}
\caption{Reconstructed anatomical B-mode images, M-mode images, and line profiles (a-e), as well as Color-Doppler images (f-j) using either uniform (a, b, f, g), or learned sparse array designs by DPS (c, d, h, i) using a sub-sampling factor of 4.  For both sampling strategies, we compare the jointly learned task model $\gtheta$ (b, g, d, i) to a naive non-learned method (DAS beamforming in a and c, followed by the Kasai auto-correlator in f and h). Fully sampled images are shown as a reference in (e, j). Dashed lines in the top and bottom image indicate the selected radial M-mode line and B-mode frame, respectively, and the red arrow indicates the depth for the displayed line profiles.}
\label{fig:ChannelResults}
\end{figure*}
\subsection{Channel sub-sampling in ultrasound imaging}
\label{ScenarioChannels}

\noindent \Cref{fig:ChannelResults} displays the anatomical B-mode and Doppler reconstructions for learned channel sub-sampling by DPS, compared to uniform channel sub-sampling. Interestingly, for both tasks, the abundant sampling of center channels was prioritized by DPS.

Assessing the respective contributions of sampling and subsequent learned processing to the overall performance of these sparse arrays, we find that using adequate jointly learned processing (here in the form of a deep network) is crucial (\cref{fig:ChannelResults}d and i). Compared to naive DAS/Kasai processing (\cref{fig:ChannelResults}c and h), we indeed observe boosted lateral resolution, contrast, and mitigation of aliasing artifacts. Yet, despite the power of these learned deep networks, they do not recover the severely aliased signals obtained by the (non-learned) uniformly sampled sparse array - strong grating lobes dominate the image in that case (\cref{fig:ChannelResults}b and g).

For phased array US probes, it is common practice to adopt a pitch which is half the signal's wavelength in order to prevent grating lobes in the \hl{field} of view \cite{szabo2004diagnostic}. Increasing the pitch between channels by uniformly sub-sampling the channel array thus caused grating lobes to appear in the B-mode images, both in grayscale (\cref{fig:ChannelResults}a and b) and with Doppler overlay (\cref{fig:ChannelResults}f and g), which we indicated by the white dashed lines in \cref{fig:ChannelResults}a and f-top. For B-mode recovery, we also show a slice of the provided B-mode frames (middle plots in \cref{fig:ChannelResults}a-e) at an axial depth indicated by the red arrow in \cref{fig:ChannelResults}e. The relative angle $\zeta_g$ of the $k^\text{th}$ grating lobes (with respect to the main beam) can be calculated as \cite{szabo2004diagnostic}:
\begin{equation}
    \label{eqn:gratingLobes}
    \zeta_g = \pm \arcsin(\frac{k\lambda}{\Delta x \cdot \frac{N}{M}}),
\end{equation}
where $\Delta x$ ($=0.151~$mm) is the (original) pitch of the array, and $\lambda$ ($=0.3~$mm) is the wavelength of the signal. For $\frac{N}{M}=4$, the angle of the first ($k=1$) grating lobe is $\zeta_g=29.8 \degree$, which is in accordance with the visually present grating lobes. 

\begin{figure}[t]
\includegraphics[width=\linewidth,trim={0cm 20.5cm 9.4cm 1.4cm},clip]{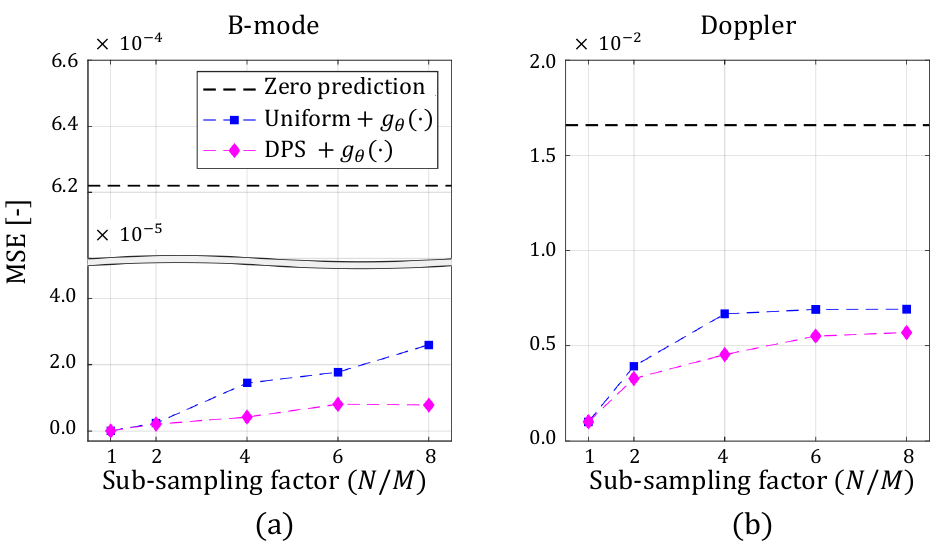}
\caption{MSE for B-mode (a) and Doppler (b) reconstruction after channel sub-sampling, averaged over the hold-out test set. The black dashed lines indicate the MSE in case only zero values are predicted. }
\label{fig:MSEgraphCH}
\end{figure}

\Cref{fig:MSEgraphCH}a and b show quantitative results for both B-mode and Doppler reconstruction using different sub-sampling factors and the two different sampling strategies (uniform versus DPS), while adopting the dedicated trained task model for reconstruction. Here we again observe that the joint training regime of sampling and reconstruction outperforms fixed sampling with a trained reconstruction model.

\begin{figure}[t]
\includegraphics[width=\linewidth,trim={0cm 19.5cm 9.3cm 2.05cm},clip]{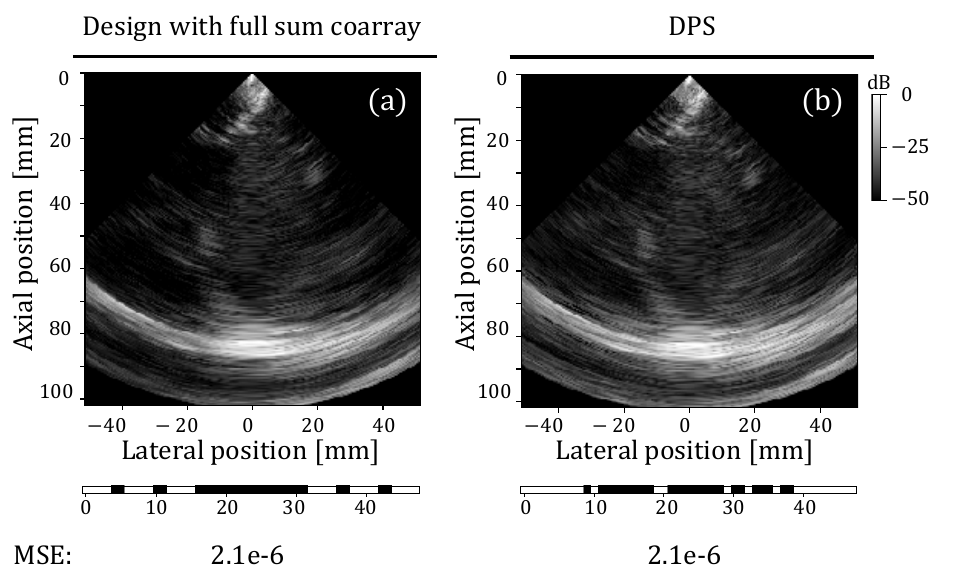}
\centering
\caption{Comparison of B-mode recovery for half of the channels selected, using a design of a sparse array that has a full sum coarray (a), therefore meeting the requirement as suggested by the authors of \cite{cohen2018beamforming}, and a realization of the learned distributions by DPS (b). Both sampling strategies are combined with a dedicated learned task model.s The reported MSE is averaged over the entire hold-out test set.}
\label{fig:CompareChannelResults}
\end{figure}

\hl{Realizing that a bigger aperture yields higher lateral resolution, it is remarkable that DPS did not assign high probability to sampling (some of the) outer channel elements of the array. %\IH{From the previously shown results we can already argue that the learned (non-linear) task model (partly) accounts for the fact that relatively many inner channels are selected.}
To check whether DPS could have found a more optimal sub-sampling pattern, possibly including more outer channel elements, we also compare DPS to a fixed sparse array (using half of the elements\footnote{The theoretical lower bound on the amount of necessary channels (allowing non-integer channel positions) is given at $2\sqrt{C}$, where the full array contains $2C-1$ elements \cite{cohen2018beamforming}. This theoretical lower bound is thus 9.9 channels for our 48-channel array. However, we can only sample integer channel positions (i.e. physically existing channels). As such, we could only create a design for sub-sampling factor $2$.}) that is designed based on the notion that its sum coarray should be full}~\cite{cohen2018beamforming}. For both sampling strategies, a dedicated task model $\gtheta$ was trained again. \hl{Notably, the learned sparse array by DPS} (\cref{fig:CompareChannelResults}b), also has a sum coarray that is full, and yields a performance that is up to par with that of the hand-designed sum coarray (\cref{fig:CompareChannelResults}). 

Computationally, running inference on patches from the test set showed an average reconstruction time of 36.7 ms ($\sigma=1.65$~ms) for IQ data from 12 channels, steered towards 68 scanlines with 2048 fast-time samples at one point in \mbox{(slow-)time}, implying a frame reconstruction rate of 27 frames per second. 

\section{Discussion}\label{Discussion}

\noindent Recent technological trends in medical imaging have spurred the demand for imaging pipelines that rely on less data without compromising image quality, temporal resolution, or more generally, diagnostics. We here consider the notion of task-adaptive sampling, in which sampling schemes are optimized not to recover the sensor signals themselves, but to fulfill a specific imaging task.

To that end, in this paper we proposed a sampling framework, DPS, that permits joint learning of a context- and task-specific sub-sampling pattern, and an adequate task model. We demonstrated that this paradigm yields improved performance compared to non-trained sampling and reconstruction models across a variety of ultrasound imaging experiments.

Unlike other recently introduced learned compressed sensing techniques, \hl{DPS} learns to sub-sample rather than to take full linear measurements, which face practical implementation challenges. Sub-sampling permits straightforward implementation of the learned sampling pattern into sensing applications, with examples being sparse array design, slow-time US pulsing schemes, (non-uniform) analog-to-digital converters (ADC) and partial Fourier measurements. 

In US imaging, we specifically applied \hl{DPS} for slow-time pulse scheme design and the array channel selection problem. Besides data reduction, the former reduces the amount of transmit events, which has the additional advantage of drastically reducing power consumption. Reduced power consumption also benefits battery life for wireless applications, and reduces heat generation of ADCs, which is particularly relevant for in-body applications. 
The applications, or tasks, that we considered within the US imaging domain were B-mode imaging and tissue-motion (color Doppler) imaging. \hl{DPS} yielded distinct sampling patterns for each task, with e.g. tissue-motion estimation spurring a pattern that uses compact groups of slow-time frames with a short inter-pulse time. 

Several points of discussion are important to note here. First, \hl{since DPS performs sampling without replacement among the distributions, duplicate sampling does not occur. Yet, in particular for channel sub-sampling, we observed that in some cases several of the $M$ distributions trained towards assigning a high probability for the same channel} (Appendix~\ref{App:probs_DPS}, \cref{fig:class_probs}b and d). The entropy penalty multiplier $\mu$ and the ratio between the learning rates $\eta_{\PHI}$ and $\eta_{\theta}$ were found to have predominant influence on this training process. Although extensive fine-tuning of hyperparameters may therefore further improve the results, it was out of the scope of this research. To promote convergence towards different distributions, one might also consider adding an additional column-wise penalty on logits matrix $\PHI$. 

Second, the experiments for channel sub-sampling were performed using a driving scheme comprising only a single diverging wave transmit, causing relatively low lateral target resolution. This may have steered the optimization of sub-sampling patterns by DPS dominantly towards the central channels. Importantly, we would like to stress that the presented sub-sampling patterns are not generic across data types or imaging setups, but rather fully context and task-adapted in an end-to-end optimized data-driven fashion. Adaptation to a new context, data distribution, and imaging setup is done based on new training data.

Third, the focus of this work was on the development of a framework that permits backpropagation-based joint learning of (discrete) sampling with a task model, and we did not heavily optimize, nor fine-tune, for different task model architectures. Comparing \cref{fig:SlowTimeresults}b and d to \cref{fig:SlowTimeresults}a and c, respectively, we notice that the convolutional nature of our task model leads to inherent denoising and loss of speckle, likely due to the deep image priors \cite{ulyanov2018deep}. Therefore, if desired, in future work, one could resort to more US-specific adaptive beamforming deep networks such as ABLE \cite{luijten2019deep,luijten2019adaptive} to better preserve speckle in B-mode images.

We expect that other US imaging applications, such as super-resolution US localization microscopy (ULM), can similarly benefit from learned and dedicated sampling schemes. In ULM, millions of highly sparse point-scatterers (intravascular microbubbles) are to be detected and localized across thousands of frames at ultrafast imaging rates~\cite{christensen2020super}. Consequently, data rates are extremely high. Recently, deep neural networks have been proposed for fast ULM recovery \cite{van2018super,van2019deep}, and one can envisage the use of \hl{DPS} to learn adequate sampling patterns that reduce data rates in this context.

Beyond the US applications considered here, future work may include learning sub-sampling and reconstruction for compressed sensing MRI \cite{lustig2007sparse}, where measurements are inherently performed by sampling the spatial Fourier domain. MRI thus shares strong similarities with signal reconstruction from partial Fourier measurements (shown in \cref{ResultsA}), making it an excellent candidate for \hl{DPS}. Also investigating the use of \hl{DPS} for sparse view CT imaging is of interest, potentially permitting reduction of the amount of transmit events, and therewith exposure to harmful radiation. 
\section{Conclusions}\label{Conclusion}
% wat laten we zien in dit paper:\
\noindent In this paper we have presented \hl{Deep Probabilistic Sub-sampling (DPS)}, a probabilistic deep learning framework that permits joint optimization of a task-based sub-sampling scheme and a downstream task model. We have demonstrated its effectiveness for sensing partial Fourier coefficients of sparse signals and a number of US imaging applications, showing that the proposed method indeed learns sampling schemes that are dedicated to a given task. As such, \hl{DPS} opens up a wide range of new opportunities; beyond US imaging, we foresee its application in other medical imaging domains (e.g. MRI and CT) and, more generally, in compressed sensing problems. 

\bibliographystyle{IEEEtran}
\bibliography{references}

\newpage
\begin{appendices}
\section{Trained probability distributions by DPS}\label{App:probs_DPS}
\begin{figure}[h]
    \includegraphics[width=\linewidth,trim={0cm 17.4cm 9.3cm 2cm},clip]{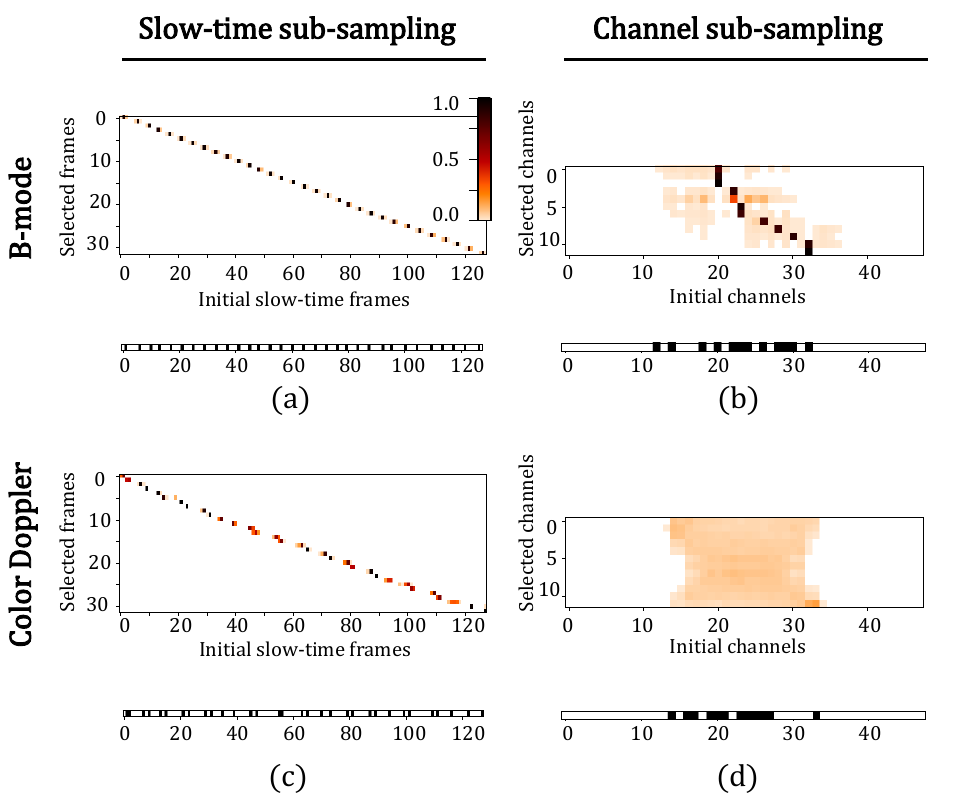}
    \caption{Learned probability distributions in DPS across ultrasound tasks (rows) and sampling strategies (columns), from which we sample without replacement to create a sub-sampling mask for selecting $M=\frac{N}{4}$ out of $N$ elements.}
    \label{fig:class_probs}
\end{figure}

\noindent DPS relies on a generative probabilistic sampling model, in which samples are selected stochastically based on learned probabilities for the entire set of candidate samples. For each selected sample, a full categorical distribution across all $N$ candidates is trained, leading to $M$ probability distributions that express the model's belief regarding the relevance of particular candidate samples at any stage of the training process. These probabilities often converged to near-one-hot representations, with high beliefs for particular candidate samples. Interestingly this did not occur for channel selection with the task of color Doppler recovery. \Cref{fig:class_probs} shows the trained probability distributions across tasks and sampling strategies, along with a particular realization of these random variables in the form of a sampling scheme. 

\end{appendices}

\end{document}